\documentclass[doublecol,figures]{epl2}
\usepackage{amssymb}
\usepackage{amsmath}
\usepackage[normalem]{ulem}

\title{Thickness-Dependent Secondary Structure Formation of Tubelike Polymers}

\author{Thomas Vogel\inst{1}\thanks{E-mail: \email{Thomas.Vogel@itp.uni-leipzig.de}}\and
Thomas Neuhaus\inst{2}\thanks{E-mail: \email{neuhaus@physik.uni-bielefeld.de}}\and
Michael Bachmann\inst{1}\thanks{E-mail: \email{Michael.Bachmann@itp.uni-leipzig.de}}\and
Wolfhard Janke\inst{1}\thanks{E-mail: \email{Wolfhard.Janke@itp.uni-leipzig.de}, Homepage: www.physik.uni-leipzig.de/CQT.html}}
\shortauthor{T.\ Vogel, T.\ Neuhaus, M.\ Bachmann and W.\ Janke}

\institute{
\inst{1} Institut f\"ur Theoretische Physik and Centre for Theoretical Sciences (NTZ), 
Universit\"at Leipzig, Postfach 100\,920, D-04009 Leipzig, Germany\\
\inst{2} John von Neumann Institute for Computing (NIC), Forschungszentrum J\"ulich, D-52425 
J\"ulich, Germany}


\abstract{By means of sophisticated Monte Carlo methods, we investigate the conformational 
phase diagram of a~simple model for flexible polymers with explicit
thickness.  The thickness constraint, which is introduced
geometrically via the global radius of curvature of a~polymer
conformation, accounts for the excluded volume of the polymer and
induces cooperative effects supporting the formation of secondary
structures.  In our detailed analysis of the temperature and thickness
dependence of the conformational behavior for classes of short
tubelike polymers, we find that known secondary-structure segments
like helices and turns, but also ringlike conformations and stiff rods
are dominant intrinsic topologies governing the phase behavior of such
cooperative tubelike objects. This shows that the thickness constraint
is indeed a~fundamental physical parameter that allows for 
a~classification of generic polymer structures.}

\begin{document}

\maketitle

\section{Introduction}
Resolving structural properties of single molecules is a~fundamental issue as
molecular functionality strongly depends on the capability of the molecules
to form stable conformations. Experimentally, the identification of substructures 
is typically performed, for example, by means of single-molecule microscopy, X-ray 
analyses of polymer crystals, or NMR for polymers in solution. With these methods,
structural details of \emph{specific} molecules are identified, but these
can frequently not be generalized systematically with respect to 
characteristic features being equally relevant for different polymers.

Therefore, the identification
of generic conformational properties of polymer classes is highly desirable. The to-date
most promising approach to attack this problem is to analyze
polymer conformations by means of comparative computer simulations of polymer models on 
mesoscopic scales,
i.e., by introducing relevant cooperative degrees of freedom and additional constraints.
In a~typical modeling approach of this kind, the linear 
polymer is considered as a~chain of beads and springs, where the monomeric properties, e.g., 
caused by the side chain,
are accumulated in an effective, specifically parametrized single interaction point of
dimension zero (``united atom approach''). The linear extension of the chain is 
accounted for by introducing springs or
stiff bonds to mimic covalent bonds in an effective way. Noncovalent van der Waals interactions between
pairs of monomers are typically modeled by Lennard-Jones (LJ) potentials. In such models, 
only the repulsive short-range part of the LJ potentials keeps pairs of monomers apart. Such
models have proven to be quite useful in identifying universal aspects of global 
structure formation processes. Examples include the characterization
of folding channels known from natural 
proteins~\cite{ssbj1} and coupled binding-folding aggregation phenomena~\cite{jbj1}, but are
by no means limited to this specific sort of polymers. 

For the identification of underlying \emph{secondary} structure
segments like helices, strands, and turns as ground states,
however, the modeling of volume exclusion by means of pure
LJ pair potentials is not sufficient to form
clearly distinct secondary structures enabling a classification scheme. 
Segments of such secondary structures  
were found, e.g., in dynamical LJ polymer studies of transient states occurring in the collapse
process~\cite{sabeur_schmid} or as ground states
in models with stiffness~\cite{noguchi}, explicit hydrogen
bonding~\cite{hoang,wolff}, or explicit solvent
particles~\cite{snir,hansen-goos_dietrich}. It could also be shown that
helical structures form by introducing anisotropic monomer--monomer potentials
in conjunction with a wormlike backbone model~\cite{kemp1} or by combining
excluded volume and torsional interactions~\cite{rapa1}.

The formation of secondary structures requires
cooperative behavior of adjacent monomers, i.e., in addition to
pairwise repulsion, information about the relative position of the
monomers to each other in the chain is necessary to effectively model
the competition between noncovalent monomeric attraction and
short-range repulsion due to volume exclusion
effects~\cite{bana1}. The simplest way to achieve this in a~general,
mesoscopic model is to introduce a~hard single-parameter thickness
constraint and, thus, to consider a~polymer chain rather as
a~three-dimensional tubelike object than as a~one-dimensional,
linelike string of monomers~\cite{bana2,bana3}. Note that this
approach differs significantly from frequently studied cylindrical
tube models~\cite{auer1}, where the tube thickness only mimics volume
exclusion but not cooperativity such that explicit modeling of
hydrogen bonds is required to generate secondary structures.

\section{Model and Methods}

In this Letter, we analyze the general thermodynamic (pseudo)phase
diagram of secondary polymer structures in dependence of the thickness
constraint.  The thickness will, therefore, be considered as coupling
parameter that separates the different conformational phases polymers
generally can reside in.  A~natural choice for parametrizing the
thickness of a~polymer conformation with $N$ monomers,
$\textbf{X}=(\textbf{x}_1,\ldots,\textbf{x}_N)$, is the global radius
of curvature $r_\text{gc}$~\cite{gonz1}. It is defined as the radius
$r_\text{c}$ of the smallest circle connecting any three different
monomer positions $\textbf{x}_i$, $\textbf{x}_j$, $\textbf{x}_k$
($i,j,k=1,\ldots,N$):
\begin{equation}
\label{eq:grc}
r_\text{gc}(\textbf{X})=\min\{r_\text{c}(\textbf{x}_i,\textbf{x}_j,\textbf{x}_k)\;\forall i,j,k\,|\,i\neq j\neq k\}.
\end{equation}
Denoting the distance between two points by
$r_{ij}=|\textbf{x}_i-\textbf{x}_j|$ and the area of the triangle
spanned by any three points by
$A_\Delta(\textbf{x}_i,\textbf{x}_j,\textbf{x}_k)$, $r_\text{c}$ is
given as
\begin{equation}
\label{eq:lrc}
r_\text{c}=\frac{r_{ij}r_{jk}r_{ik}}{4A_\Delta(\textbf{x}_i,\textbf{x}_j,\textbf{x}_k)}.
\end{equation}
With these definitions, the polymer tube $\textbf{X}$ has the
``thickness'' (or diameter) $d(\textbf{X})=2r_\text{gc}(\textbf{X})$
which is illustrated in an intuitive 
way in Refs.~\cite{gonz1,neuhaus1}.

We here consider linear, flexible polymers with stiff bonds of unit
length ($r_{i\,i+1}=1$).  The pairwise interactions among nonbonded
monomers are modeled by a~standard LJ potential and thus the energy of
a~conformation $\textbf{X}$ reads
\begin{equation}
\label{eq:lj}
E(\textbf{X})=\sum\limits_{i,j>i+1} V_{\mathrm{LJ}}(r_{ij}),
\end{equation}
where $V_{\mathrm{LJ}}(r_{ij})=4\varepsilon[(\sigma/r_{ij})^{12}-(\sigma/r_{ij})^6]$.
By setting $\sigma=1$, $V_{\mathrm{LJ}}(r_{ij})$ vanishes for $r_{ij}=1$ and is minimal at
$r_{ij}^{\mathrm{min}}=2^{1/6}\approx 1.122$.

Since we are interested in classifying conformational pseudophases of polymers
with respect to their thickness, we introduce
the restricted conformational space ${\cal R}_\rho=\{\textbf{X}\,|\, r_{\mathrm{gc}}(\textbf{X}) > \rho\}$
of all conformations $\textbf{X}$ with a~global radius of curvature 
larger than a thickness constraint $\rho$, which can be understood
as an effective measure for the extension of the polymer side chain. Given $\rho$, obviously
only conformations with $r_\text{gc}\ge \rho$ can occur.

The canonical partition function of the restricted conformational space thus reads
\begin{equation}
\label{eq:partsum}
Z_\rho=
\int {\cal D}X\,\Theta(r_\text{gc}(\textbf{X})-\rho)e^{-E(\textbf{X})/k_\mathrm{B}T},
\end{equation}
where $k_\mathrm{B}T$ is the thermal energy (we use units in which 
$\varepsilon=k_\mathrm{B}=1$ in the following) and $\Theta(z)$
is the Heaviside function. In this thickness-restricted space, canonical
statistical averages of any quantity $O$ are then calculated via 
$\langle O\rangle_\rho=Z_\rho^{-1}\int {\cal D}X\,O(\textbf{X})
\Theta(r_\text{gc}(\textbf{X})-\rho)\exp[-E(\textbf{X})/k_\mathrm{B}T]$.

To characterize the phase diagram, we have first performed exhaustive
energy-landscape paving (ELP) optimizations~\cite{hansmann1} in order to identify 
lowest-energy conformations as reference states of flexible polymers 
under the constraint of a~given minimal global radius of
curvature~$\rho$. Next, we have analyzed the interplay of structural and thermal
properties of polymers with $N=8,\ldots,13$ monomers for a~very large number of 
$\rho$ values.
It should be emphasized that our detailed thermodynamic analysis aiming at the entire
structural phase diagram requires precise datasets
that can only be obtained by means of sophisticated generalized-ensemble methods.
We have employed parallel tempering~\cite{pt1}, multicanonical
sampling~\cite{muca1}, and the Wang-Landau method~\cite{wl1}, and compared the results. 
In the following, we 
shall focus on the tube polymer with $N=9$ monomers as most of the observed features 
of this 9mer are 
generic and thus also common to the longer chains. 

\section{Results and Discussion}

\begin{figure}
\onefigure[width=\columnwidth]{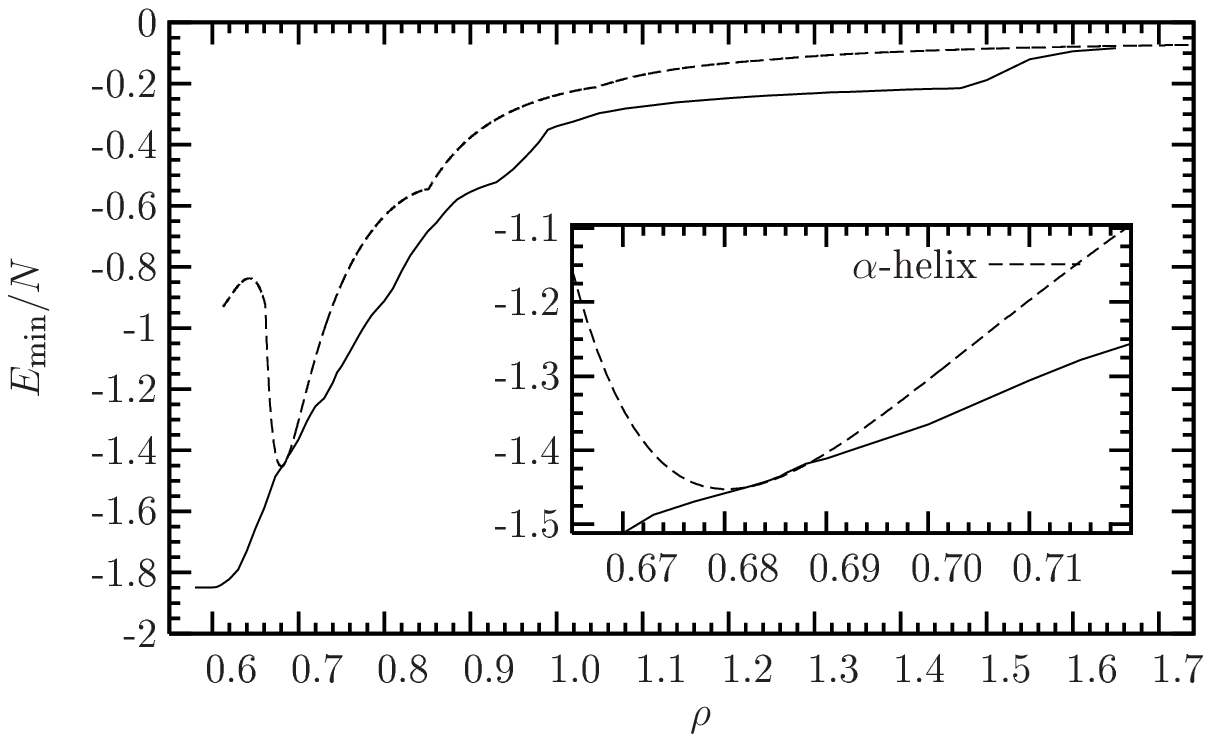}
\onefigure[width=\columnwidth]{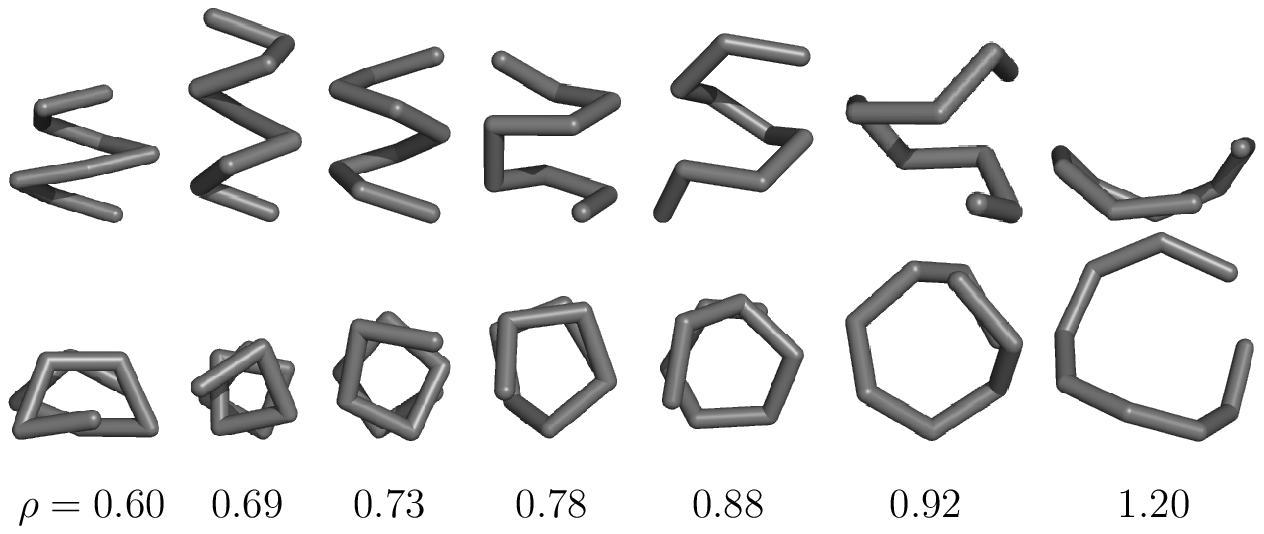}
\caption{Ground-state energy per monomer $E_\text{min}/N$ of tubelike polymers with nine monomers as a~function of 
the global radius of curvature constraint~$\rho$ (solid line). For comparison, also the energy curve
of the perfect $\alpha$-helix is plotted (dashed line). The inset shows that for a~small
interval around $\rho\approx 0.686$, the ground-state structure  
is perfectly $\alpha$-helical. Also depicted are
side and top views of
putative ground-state conformations for various exemplified values of~$\rho$.
For the purpose of
clarity, the conformations are not shown with their natural thickness.}
\label{fig:minE}
\end{figure}

Figure~\ref{fig:minE} shows the ground-state energy per monomer as
a~function of~$\rho$ (with bin sizes $\Delta \rho\le 0.01$ in the most
interesting region). Also shown are lowest-energy conformations for
exemplified values of~$\rho$.  The ground-state energy per monomer for
the linelike 9mer (i.e., $\rho=0$) is in our units
$E_\text{min}/N=-1.85$. The thickness constraint becomes relevant, if
$\rho$ is larger than half the characteristic length scale
$r_{ij}^{\mathrm{min}}$ of the LJ potential: In the interval
$2^{-5/6}\approx 0.561 < \rho < \rho_\alpha \approx 0.686$
conformations are pre-helical.  The nonbonded interaction distance is
still allowed to be so small that structures are
deformed. Nonetheless, the onset of helix formation is clearly visible
as it is an intrinsic geometrical property of any linelike
object. Optimal space-filling helical symmetry is reached when
approaching $\rho_\alpha$, where the ground-state conformation takes
the perfect $\alpha$-helical shape (see inset of Fig.~\ref{fig:minE}).
All torsional angles are identical (near $41.6^\circ$) and also all
local radii are constant; the number of monomers per winding is
3.6. Note that for proteins, where the effective distance between two
C$^\alpha$ atoms is about 3.8\,{\AA}, $\rho_\alpha$ in our units
corresponds indeed to a~pitch of about 5.4\,{\AA} as known from
$\alpha$-helices of proteins.  Thus, an $\alpha$-helix is a~natural
geometric shape for tubelike polymers. Hydrogen bonds stabilize these
structures in nature---but are not a~necessary prerequisite for
forming such secondary structures.

For larger values of~$\rho$, helices unwind, i.e., the pitch gets larger and the number of monomers 
per winding increases. However, helical structures still dominate the ground-state 
conformations. It should be noted that our model is energetically invariant under helicity reversal,
i.e., left-handed helices or segments are not explicitly disfavored and are, therefore, also 
equally present 
in the conformational space. In the interval $\rho_\alpha\le \rho \lesssim 0.92$, fluctuation peaks of the 
derivative $\upd E_\text{min}/\upd\rho$ (not shown) indicate that there are also stable helical conformations
in the vicinity of $\rho \approx 0.73$ (winding number $\approx 4.5$) and $\rho \approx 0.78$ 
(winding number $\approx 5.0$). Near $\rho \approx 0.92$, the final helical state has been reached. 
The thickness 
has increased in such a~way that the most compact conformation is a~helix with a~single winding. 
After that, a~topological change occurs and the ground-state conformations are getting
flatter. The helix finally opens up and planar conformations with similarities to $\beta$-hairpins
become dominant. These structures are still stabilized by nonbonded LJ interactions
between pairs of monomers. Increasing the thickness further leads to a~breaking of these contacts
and ringlike conformations become relevant~\cite{neuhaus1}. We have verified that for values
$r_{\mathrm{gc}}\approx N/2\pi$, ground-state conformations are almost perfect circles 
with radius $r_{\mathrm{gc}}$. The existence of ringlike conformations is a~consequence of the long-range
monomer-monomer attraction. 
Eventually, for $\rho\to\infty$, the effective stiffness increases, also the end contacts 
disappear, and only thick rods are still present. 

After these preparatory considerations of ground-state properties, we are now going to discuss 
the thermodynamic behavior
of the tube polymers. Based on the peak structure of the specific heat as a~function
of temperature~$T$ and thickness constraint~$\rho$, we identify the
structure of the conformational $\rho$-$T$ pseudophase diagram. 
We do this again for the 9mer which allows for a~very precise analysis. Only for such  
a~small system, hundreds of separate generalized-ensemble computer simulations can be performed. However,
we verified the results also for larger polymers with up to 13 monomers and found that there are
no significant changes in the phase-diagram topology~\cite{vbnj1}. 
Even the expected shifts of the transition lines due to finite-length corrections 
are very small such that we have good reason to assume that the pseudophase diagram of the 9mer
reflects the
general phase structure of short tubelike polymers pretty well. 
This is partly due to the fact that the polymer thickness as defined via the global radius
of curvature is a~length-independent constraint and the chains in our study are short enough to prevent
the formation of tertiary structures (as, e.g., arrangements of different secondary-structure
segments forming a~tertiary domain). For longer chains, however, tertiary 
structures are definitely relevant. The longest chain in our study, the polymer with $N=13$ monomers,
already exhibits first indications of structure formation on globular length scales.
However, the description 
of such tertiary folding processes is not in the focus of the present work.
Compact globular conformations form by decreasing the temperature below
the $\Theta$ point whose properties for tubelike polymers are
a~study worth in its own right.

\begin{figure}
\onefigure[width=\columnwidth]{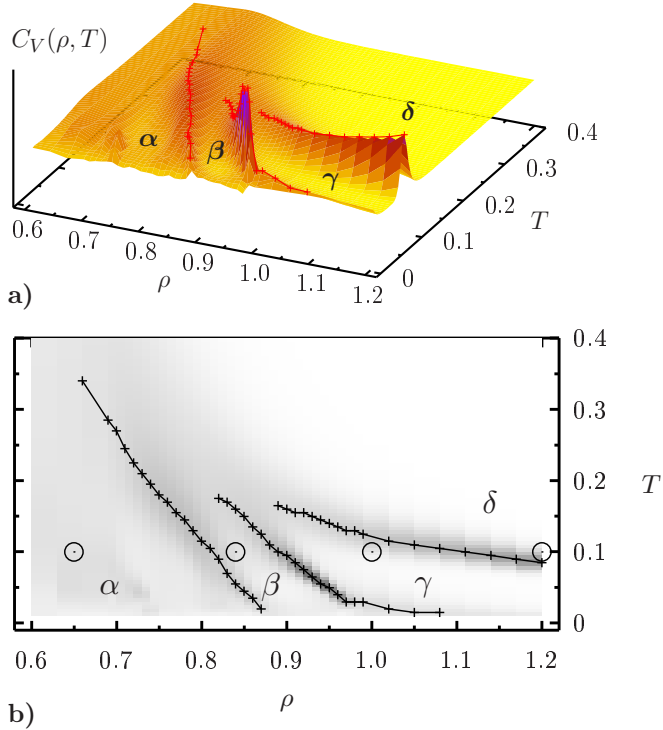}
\caption{
(a) Perspective and (b) projected
view of the specific-heat profile $C_V(\rho,T)$ for a 9mer which is interpreted
as structural pseudophase diagram of thermodynamically relevant 
tube polymer conformations in thickness--temperature parameter space.
Dark regions and data points ($+$)
indicate the ridges of the landscape and
separate conformational phases.} Helical or helix-like conformations
dominate in region~$\alpha$, sheets in region~$\beta$, rings in
region~$\gamma$, and stiff rods in pseudophase~$\delta$. Circles
($\odot$) indicate the locations where the exemplified conformations
of Table~\ref{tab:confs} are relevant. The general structure of the
phase diagram remains unchanged also for the longer polymers
considered in our study.
\label{fig:pd}
\end{figure}

Our main results are contained in the phase diagram of Fig.~\ref{fig:pd}, which 
shows the specific-heat landscape 
$C_V(\rho,T)=(\langle E^2\rangle_\rho-\langle E\rangle_\rho^2)/T^2$ for a 9mer
as obtained from reweighting the density of states for given thickness constraint~$\rho$. 
Dark regions correspond to strong energetic fluctuations, i.e., the 
darker the region the larger is the
specific-heat value. Data points ($+$) mark the peaks or ridges of the profile and 
indicate conformational activity and thus represent transitions between different conformational
pseudophases. Error bars  
are not shown for clarity but 
are sufficiently small (for most data points smaller than symbol size), 
so that the identified pseudophase boundaries are statistically significant. 

Guided by the analyses of the ground-state properties, we identify
four principal pseudophases\footnote{We note that there are singular points in the 
parameter space corresponding to special geometric representations of secondary structures.
For the chain with length $N=8$ and 
$r_{\mathrm{gc}}\approx 1/\sqrt{2}$, for example, 
the degenerate ground-state conformation exhibits an almost perfect alignment of 
the chain along the edges of a~cube.}. In region $\alpha$, helical conformations  
are the most relevant structures. In particular, the $\alpha$-helix
resides in this pseudophase. Characteristic for the transition 
from pseudophase $\alpha$ to $\beta$ is the unwinding of the helical structures which are getting
more planar. Thus, region $\beta$ is dominated by simple sheet-like structures. Since the 9mer
is rather short, the only sheet-like class of conformations is the hairpin. For longer chains, one 
also finds more complex sheets, e.g., lamellar structures~\cite{bana2,vbnj1}. 
A~characteristic property of the
hairpins is that these are still stabilized by nonbonded interactions. These break with 
larger thickness and higher temperature. Entering pseudophase $\gamma$, dominating structures
possess ringlike shapes. Finally, region $\delta$ is the phase of random coils, which are 
getting stiffer for large thickness and eventually resembling rods. 
Representative polymer conformations dominating
the pseudophases in the regions $\alpha$ to $\delta$ are depicted in Table~\ref{tab:confs} in different 
representations. 

\begin{table}
\caption{Exemplified conformations being thermodynamically relevant 
in the respective pseudophases shown in Fig.~\ref{fig:pd}, visualized in different representations.}
\label{tab:confs}
\begin{center}
\begin{tabular}{ccc}\hline
phase & type & views of representative example  \\ \hline\\[-10pt]
$\alpha$ & helix & 
\parbox{5.9cm}{\centerline{\onefigure[width=5.9cm]{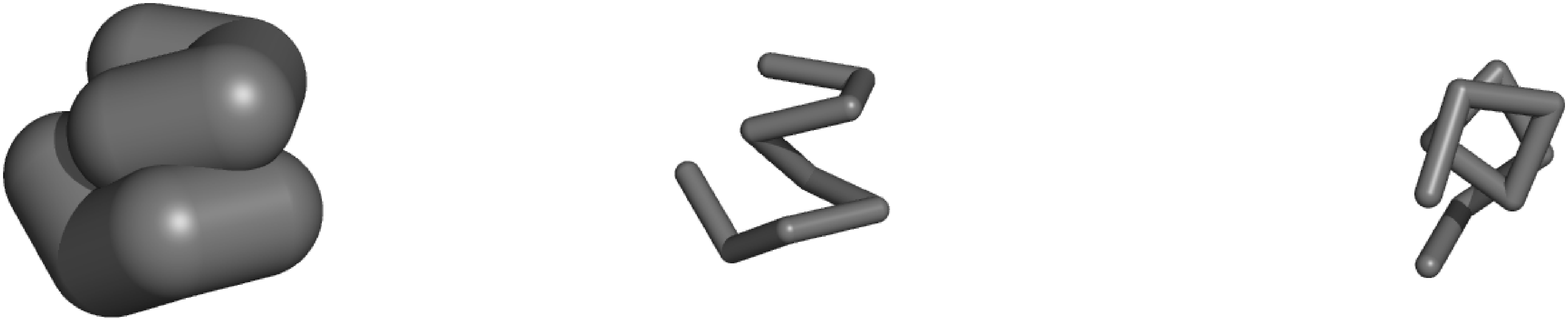}}}\\
$\beta$ & sheet &
\parbox{5.9cm}{\centerline{\onefigure[width=5.9cm]{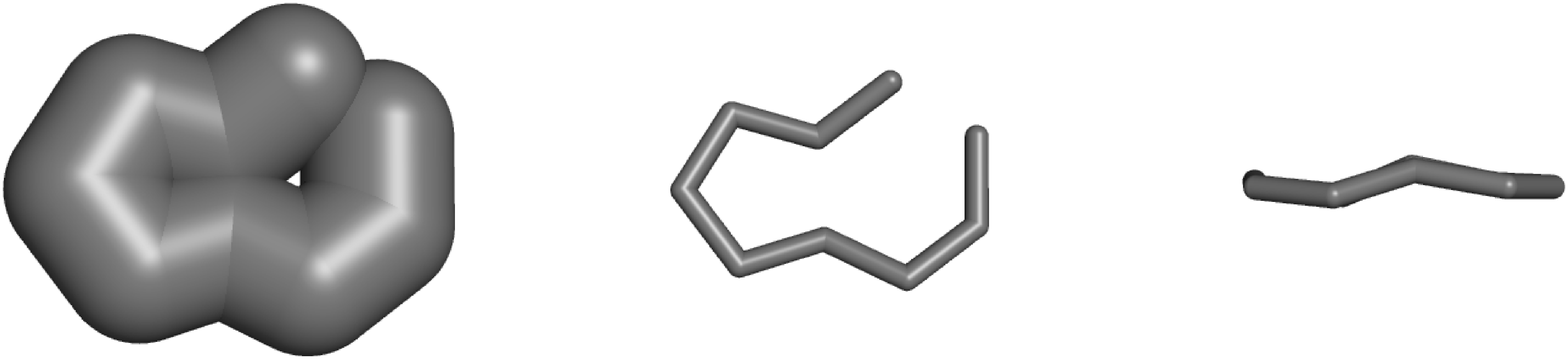}}}\\
$\gamma$ & ring &
\parbox{5.9cm}{\centerline{\onefigure[width=5.9cm]{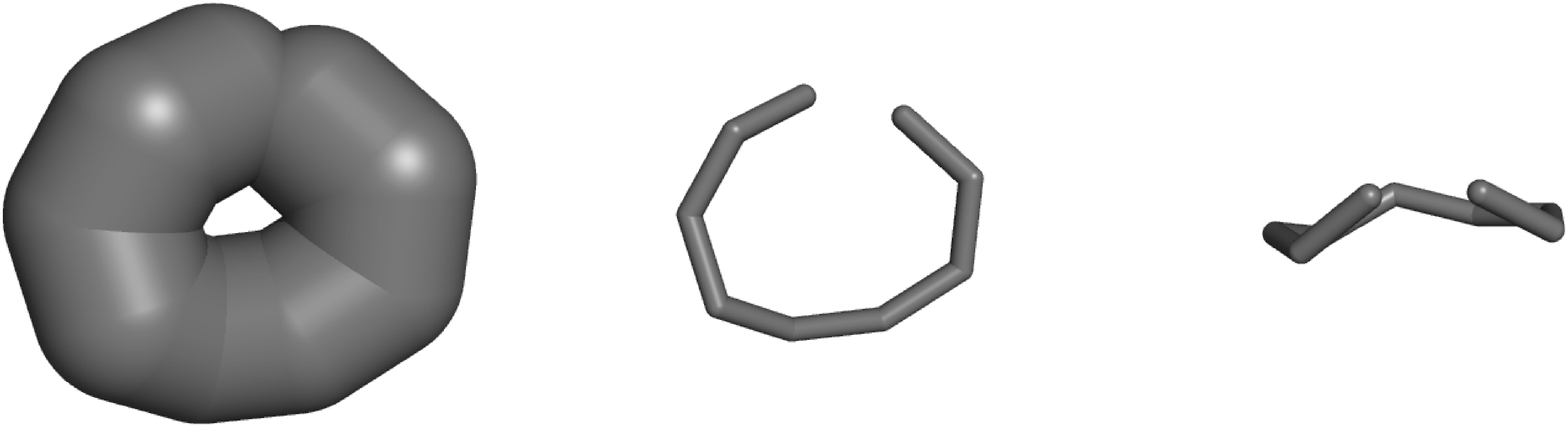}}}\\
$\delta$ & rod & 
\parbox{5.9cm}{\centerline{\onefigure[width=5.9cm]{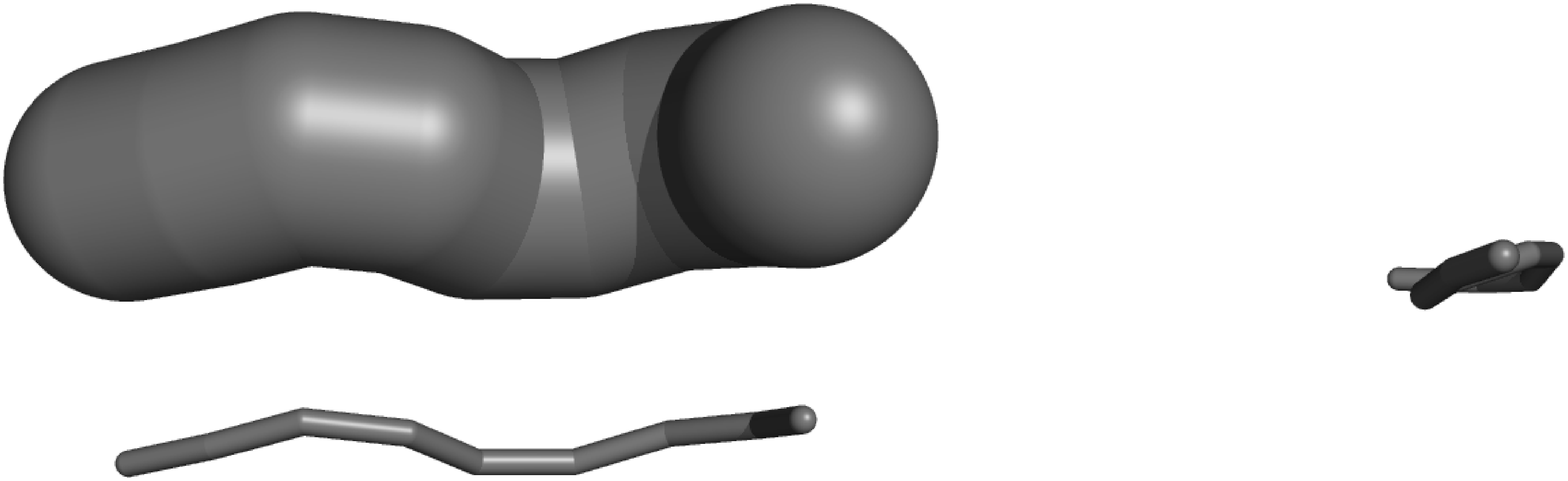}}}\\ \hline
\end{tabular}
\end{center}
\end{table}

\section{Summary}
In this Letter we have focused on an analysis of the thermodynamic
properties of tubelike polymers. The tube picture is a~simplification 
of the volume extension of polymers due to steric constraints of their
backbone or the presence of side chains. The thickness of such a~mesoscopic
tube can be considered as a~single steric parameter that induces cooperative effects
and permits the discrimination of
polymers. Thus, the phase diagram presented here does not only allow for the classification
of possible thermodynamic conformational phases of a~single polymer with fixed thickness. 
Rather, performing generalized-ensemble 
simulations for different thicknesses enabled us for the first time to 
resolve the complete  
(pseudo)phase behavior with respect to the thickness constraint and temperature. 
This means that the we have
identified the generic structure of the conformational phase space at non-zero temperatures
for \emph{classes} of polymers,
parametrized by their thickness. Although we employ a~mesoscopic model for flexible polymers,
we find that the thickness constraint is an intrinsic source of an effective stiffness
and enhances 
the capability of a~polymer to form secondary structures which are stable against
thermal fluctuations. The stability limits for increasing temperature are elucidated
in the phase diagram in Fig.~\ref{fig:pd} which summarizes our main findings.
In particular, we clearly find 
helical and sheet-like structures which are dominant in different pseudophases.
Thus, the thickness is indeed 
a~fundamental physical parameter that
allows for the classification of polymers with respect to their transition behavior
and their preference to form characteristic secondary structures, depending on external
parameters such as temperature.

Since the development of high-resolution experimental techniques is breathtakingly 
advancing, the interest in structural properties on nanoscopic scales is increasing,
in particular when it comes to applications where small molecules are used as building 
blocks in the design of functional molecular machines. In these cases the understanding of
the effect of sterically induced constraints on molecular structure formation is of 
particular importance.

\acknowledgments
This work is partially supported by the 
DFG (German Science Foundation) under Grant
Nos.\ JA 483/24-1/2 and the Graduate School of Excellence ``BuildMoNa''.
Some simulations were performed on the
supercomputer JUMP of the John von Neumann Institute for Computing (NIC), Forschungszentrum
J\"ulich, under Grant No.\ hlz11.

\end{document}